\documentclass[aps,pre,superscriptaddress,preprintnumbers,twocolumn,
amsmath,amssymb,floatfix]{revtex4} \usepackage{graphicx}
\usepackage{bm}
\begin{document}

\author{Roya Zandi}
\affiliation{Department of Chemistry and Biochemistry, UCLA,
Box 951569, Los Angeles, California, 90095-1569}
\affiliation{Department of Physics, Massachusetts Institute of
Technology, Cambridge, MA 02139, USA}

\author{Yacov Kantor} \affiliation{School of Physics and Astronomy,
Raymond and Beverly Sackler Faculty of Exact Sciences, Tel Aviv
University, Tel Aviv 69978, Israel}

\author{Mehran Kardar}
\affiliation{Department of Physics, Massachusetts Institute of
Technology, Cambridge, MA 02139, USA}

\date{\today}

\title{Entropic competition between knots and slip-links}

\begin{abstract}

Using canonical Monte Carlo simulations, we introduce a new
numerical procedure for comparing the entropic exponents of
polymers with different constraints and/or topologies.  Setting up
competitions between polymer segments which can exchange monomers
according to their free energies, we obtain the universal
exponents of partition functions, independently of any knowledge
of the non-universal part.  The method is successfully tested for
closed polymer loops decorated with sliding rings.  We also
investigate the entropic exponents of loops with a fixed knot
type, in which case we are limited by strong finite--size effects.

\end{abstract}

\maketitle

\section{Introduction and Summary} \label{sec:intro}
A long flexible polymer chain can assume an enormous number of
configurations; thus, the conformations of polymers are best described
statistically.  Despite the microscopic distinctions which exist among
different polymers, at large length scales many generic properties are
independent of the details of specific polymer structure and thus are
``universal.''\cite{degennes} The understanding and characterization
of this universal behavior has led to great advances in the
statistical treatment of polymers.  Polymers in good solvents have
been modeled by self-avoiding walks (SAWs), which are in turn mapped
to a magnetic system at its critical point \cite{degennes}. 
Renormalization group techniques then allow a variety of polymer
properties to be calculated analytically\cite{degennes,freed}.  On the
experimental front, advances in manipulation and imaging of single
molecules have provided new impetus for studies of polymers
\cite{smith}.  With these micromanipulation techniques, it is now
possible to explore a wide variety of physical properties of polymers,
and thus to test models and theoretical predictions at the level of a
single chain.

In many circumstances, the number of configurations of a polymer grows
with its length $\ell$, as
\begin{equation}
w (\ell) \simeq
A\mu^{\ell}\ell^{-c}\left[1+B\ell^{-\Delta}+\cdots\right].  \label{w0}
\end{equation}
In the above equation, the connectivity constant $\mu$, is a
non-universal quantity which depends on the microscopic features of
the chain, but (like a free energy density) remains the same for
different boundary and topological constraints\cite{degennes}.  By
contrast, $c$ is a {\em universal} exponent, which is independent of
the microscopic characteristics of the polymer, but which does depend
on the dimensionality, and on global boundary and topological
constraints.  For example, in the case of a closed loop, $c=d\nu$,
where $d$ is the dimensionality of space, and $\nu$ is the exponent
relating the typical spatial extent of a polymer to its length by
$R\sim \ell^\nu$.  (In $d=3$, for polymers with self-avoiding
interactions $\nu\approx 0.588$\cite{degennes, khokhlov}.)  In
Eq.~(\ref{w0}), we have anticipated subleading corrections to the
leading asymptotic behavior at large $\ell$, indicated by so-called
corrections to scaling in the square brackets.  Renormalization group
calculations suggest that the exponent $\Delta$ is universal, while
the amplitude $B$ is case-specific\cite{freed}.

Numerical determination of a power-law correction to the leading
exponential behavior is rather cumbersome.  It is usually accomplished
by examining a generating function $f(z)=\sum_{\ell} \ell^g
w(\ell)z^\ell$, with a conveniently chosen value of the free parameter
$g$.  The function $f(z)$ is usually singular at $z=1/\mu$
\cite{madras}, and the details of this singularity determine the
exponent $c$ appearing in Eq.  \ref{w0}.  This can be done either by
examining a finite amount of terms in series expansions, using
$w(\ell)$ extracted from exact enumeration studies, or by performing
grand-canonical Monte Carlo(MC) studies of the polymer with weight
determined by $\ell^g z^\ell$ for several values of $z$, as was done
in Refs.  \cite{orlandini,guitter}.

Here, we propose a method for calculating the universal exponent $c$
in the number of configurations, without the need to account for the
leading non-universal exponential growth $\mu^\ell$.  Our procedure is
best suited for the evaluation of the exponent $c$ when the polymer is
constrained in various manners, a situation that occurs physically
when a polymer can move from one regime to another.  Using canonical
Monte Carlo simulations, we directly compare the {\em relative} number
of configurations by setting up an ensemble in which two segments of a
polymer can exchange monomers.  A pair of monomers belonging to two
different segments does not interact, while the monomers in the same
segment repel each other.  Each segment is subject to its own set of
global constraints.  The total number of monomers is fixed to $L$,
while in any given configuration the two segments have $\ell$ and
$L-\ell$ monomers, respectively.  According to Eq.~(\ref{w0}), in the
asymptotic limit, the number of configurations of such a system is
\begin{equation}
w(\ell,L)=A\ell^{-c_1}(L-\ell)^{-c_2}, \label{length}
\end{equation}
where $c_1$ and $c_2$ are the exponents characterizing each
segment; the prefactor $A$ is independent of $\ell$. Assuming that the
asymptotic limit has been reached, fitting the histogram of $\ell$
should provide the exponents $c_1$ and $c_2$.

\begin{figure}
\centerline{\includegraphics[height=3.0in]{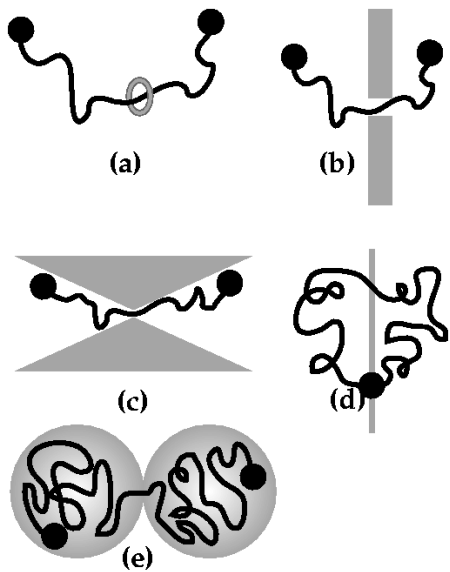}}
\caption{Examples of possible applications of ``entropic competition:''  (The
universal power-law exponents depend on the boundary condition of the
system.)  (a) is appropriate for calculation of the exponent $\gamma$ which
describes open linear polymers.  Note that the two segments of the
chain do not interact with each other.  (b) is good for calculation of the
exponent $\gamma_{1}$, (c) for a wedge, (d) for exponent
$\gamma_{11}$, and (e) for a chain attached at one end to the 
interior surface of a sphere.}
\label{fig:all}
\end{figure}

This method, which we call ``entropic competition,'' can in
principle be applied to several physical circumstances, some of
which are depicted in Fig.~\ref{fig:all}.  For example, one can
extract the entropic exponents of linear chains by threading them
through a hole, and making the first and last monomers of the
chain bigger than the size of the hole such that the chain can
diffuse back and forth without wandering away
(Fig.~\ref{fig:all}a).  A key point in this set-up is that the two
parts of the chain are not allowed to interact with each other. By
examining the probability of having a segment of size $\ell$,
which is proportional to $\ell^{\gamma-1} (L-\ell)^{\gamma-1}$,
one can extract the entropic exponent of linear chains, known as
$\gamma$ \cite{gamma}. If, instead of threading a chain through a
hole, we divide the chain into two parts by a rigid wall and let
the two parts of the chain exchange monomers at the wall (a
process called translocation), we can extract the entropic
exponent which is called $\gamma_{1}$ \cite{gamma}
(Fig.~\ref{fig:all}b).  Other potential applications are to the
calculation of entropic exponents of linear polymers in the
presence of a wedge, or inside a spherical shell
(Figs.~\ref{fig:all}(c,~e)).  The latter case is particularly
relevant to the translocation of DNA through spherical capsids and
has been the subject of intense research\cite{park}.  ``Entropic
competition'' is especially appropriate for calculation of the
entropic exponent $\gamma_{11}$, related to the polymers with two
ends anchored to a surface \cite{gamma}.  In this case, by fixing
one monomer of a ring on a wall and letting the two segments of
the ring on the two sides of the wall exchange monomers with each
other at any point on the wall, we can extract the entropic
exponents of chains with two ends restricted to move on a surface
(Fig.~\ref{fig:all}e). The latter is of renewed interest due to
potential relevance to a new class of physical gels obtained from
triblock copolymers anchored to fluid membranes at each end
\cite{safinya}.

\begin{figure}
\centerline{\includegraphics[height=1.5in]{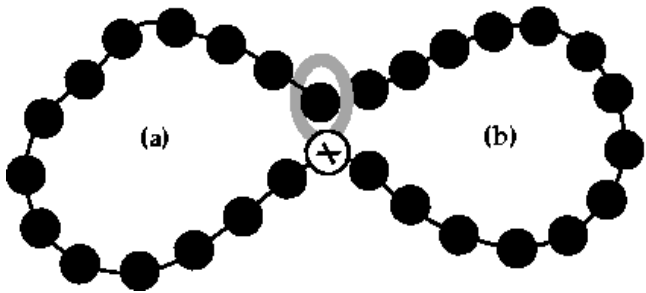}} 
\caption{A schematic
depiction of two loops in competition.  Sides (a) and (b) do not
interact with each other.  The position of one bead is always fixed
(the one with X inside).  The oval ring shows the position of the
`hole' separating the spaces in which the segments exist.}
\label{fig:fig1}
\end{figure}

The calculation of entropic exponents using ``entropic
competition'' is not solely restricted to open polymers.  One can
divide a ring into two loops with the use of a `hole' connecting
two separate spaces, and a monomer which is fixed in position, as
illustrated in Fig.~\ref{fig:fig1}.  In Sec.~\ref{sec:method} we
employ this system to calculate the probability distribution of
two {\em non-interacting} self-avoiding (SA) loops of lengths
$\ell$ and $L-\ell$, which can freely exchange monomers. If the
two polymer segments have the same topology and boundary
conditions, as in the aforementioned case, then the exponents
$c_1$ and $c_2$ of Eq.~(\ref{length}) are equal and one can
directly find the entropic exponents by fitting the probability
distributions resulting from simulations to Eq.~(\ref{length}).

The entropic exponent can in fact be modified in a systematic
manner by decorating polymer loops with sliding rings. We
investigate the effect of sliding rings on the number of
configurations of closed chains in Sec.~\ref{sec:ring}.  We assume
that the size of a ring is the same as that of one monomer.  In
this situation, the presence of a ring increases the number of
configurations of the loop by a factor of $\ell$ (the number of
monomers on the loop where the ring can slide).  Once again we
examine the accuracy of ``entropic competition'' by comparing the
results of the simulations with analytical expectations, and find
an excellent agreement between them.

Introduction of the ``entropic competition'' method was motivated
by its potential application to {\em knots}. The influence of a
knot on the entropy of polymers has been the subject of interest
for some time \cite{orlandini, guitter, janse, ralf}.  In
Sec.~\ref{sec:knot} we study knotted polymers. We overview the
current hypothesis regarding the number of configurations of a
knotted polymer. We numerically study several related models and
observe that the presence of knots changes the number of
configurations. Furthermore, we numerically calculate the
probability distribution of two loops with one trefoil knot on
each side and find that it is completely different from that of
two loops with one ring on each side.  This may well be an
indication of the importance of finite-size effects in knots, as
the two cases are conjectured to be similar in the limit of very
long chains.

Finally, in Sec.~\ref{sec:compete}, we observe the ``competition''
between a knot and rings by placing a knot on one loop and rings
on the other loop.  In our simulations, the knot pulls the chain
much harder than a single ring, possibly due to finite-size
effects. For polymers in the range of a few hundred monomers,
roughly six rings are needed to ``compete'' with a simple trefoil
knot. We attempt to quantify this finite knot-size effect.

In the absence of an exact statistical treatment of knotted polymers,
several researchers have employed `slip-links' for gaining analytical
insights on topological
constraints\cite{ralf,ball}.  Slip-links can be
envisioned as belt-buckles which force two points of the chain to be
close to each other.  Para-knots are collections of such slip-links,
and have been proposed as means of designing entropy-driven functional
molecules which are linked to each other mechanically rather than
chemically\cite{metz}.  In Sec.~\ref{sec:compete}, we also employ a
para-knot model to  explore the effects of a sub-leading scale
on entropic competition.

\section{Method} \label{sec:method}

We start by calculating the entropic exponents of a closed loop.  Our
model polymer consists of $L$ hard spheres of diameter $D$ connected
into a chain by tethers which have no additional energy cost
\cite{kantor}, but restrict the distance between connected spheres to
be smaller than $1.2 D$; this prevents the chain from crossing itself. 
Figure \ref{fig:fig1} provides a schematic depiction of the
simulation: Solid circles represent monomers of the chain; the
position of one monomer (labeled with an X) is fixed in space, and the
chain is forced to pass through a hole which is depicted by an oval
ring in Fig.~\ref{fig:fig1}.  The effect of the hole combined with the
fixed monomer is to create two loops of lengths $\ell$ and $L-\ell$. 
The two loops do not interact with each other: as soon as a monomer
passes through the hole, it ceases to interact with the monomers of
its previous side and starts to interact with the ones in the new
side.  In order to preserve the topology of each loop, we allow the
monomers immediately adjacent to the hole on either side to interact
with each other, and also allow the fixed monomer to interact with all
the other monomers.  It is also necessary to keep the hole narrow
enough not to let knots pass through it (for later applications).  The
hole creates a barrier which slows down the process of transferring
the chain from one side to the other one.  In fact, this entropic
barrier is the major obstacle to the computation of the probability
distribution of long chains ($L>300$) within a reasonable amount of
CPU time.  Obviously, this barrier depends on the entropic exponents,
and is less significant for a hole on a rigid wall.  In calculating
the exponent $\gamma_{11}$ \cite{gamma}, we remove the entropic
barrier (hole) completely; the two segments on the two sides of the
wall can exchange monomers at any point on the wall
(Fig.~\ref{fig:all}e).  In this case, we let the monomers pass through
the wall only based on their order along the chain, \emph{i.e.},
monomer $n$ can pass through the wall, only if monomer $n-1$ has
already moved to the other side of the wall.

\begin{figure}
\centerline{\includegraphics[height=1.5in]{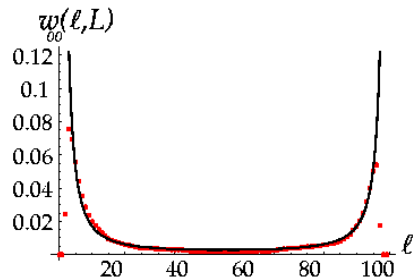}}
\caption{Probability distribution for two loops exchanging monomers.
The dots are the results of the simulation for $10^{9}$ MC steps, with
$L=100$.  The graphs are normalized such that the integrated weight is
equal to one.}
\label{fig:twoloops}
\end{figure}

The number of configurations of the system depicted in Fig.
~\ref{fig:fig1} is the product of number of configurations of each
loop, and thus scales as
\begin{equation}
w_{00}(\ell, L-\ell) = A \mu^L \ell^{-\nu d} \times (L-\ell)^{-\nu d},
\label{well}
\end{equation}
where $\ell$ is the number of monomers in one of the loops.  
(The absence of a knot or ring on each loop is denoted by the index pair $00$.)
Note that the $\mu$ does not affect the $\ell$-dependence, and shall
in fact be ignored henceforth.  Although the passage through a
worm-hole may appear unphysical in this context, our simulations
produce the correct results for the two non-interacting self-avoiding
loops which exchange monomers.  Figure~\ref{fig:twoloops} shows the
probability distribution for the segment length $\ell$.  The dots are
the result of the simulation for $10^{9}$ MC steps \cite{mcstep}.  The
solid line corresponds to the Eq.~(\ref{well}) with $\nu=0.588$ at
$d=3$.  The graphs in this figure and all the other figures are
normalized such that the integrated weight is equal to one.  The good
match of simulation with Eq.~(\ref{well}) confirms that our method for
computing the probability distribution is accurate.  Similar curves
were produced through simulations for chains with different sizes
$L=50$, $L=150$, and $L=200$; all follow Eq.~(\ref{well}).

\section{Sliding Rings}
\label{sec:ring}

To further examine the validity of our method, in this section we
present results of simulations including {\em sliding rings}.  For
each configuration of a loop of length $\ell$, a sliding ring can
occupy $\ell$ different positions, and thus the probability
distribution of such a loop scales as, $w_1 \simeq \ell w_{0} \sim
\mu^{\ell} \ell^{-\nu d +1}$.  In simulations, the passage of a
chain through the hole is stopped if a ring is placed exactly on
the monomer entering the pore.  Instead of simulating a ``real''
sliding ring, we calculate the probability that a ring might be
sitting adjacent to the hole and then prevent the chain from
passing through the hole with this probability.  If we have one
ring, the probability of the ring to be exactly on the monomer
entering the hole is $1/\ell$.  In case of many rings, this
probability goes up.

\begin{figure}
\centerline{\includegraphics[height=1.5in]{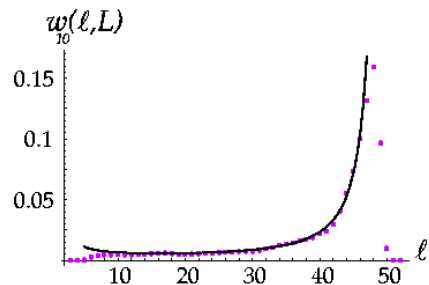}}
\caption{Probability distribution of the size $\ell$ of a loop with
one sliding ring, in competition with a simple loop.
The dots are the result of a simulation with over $10^{9}$
MC steps for $L=50$. }
\label{fig:ring1}
\end{figure}

We start by placing a ring on one of the loops of Fig.~\ref{fig:fig1}
and then we measure the histogram of lengths.  Figure~\ref{fig:ring1}
shows the probability distribution of the loop side with one sliding
ring.  The increased entropy of the segment on which the ring is
sliding biases the amount of time monomers spend on that side.  The
solid line in Fig.~\ref{fig:ring1} corresponds to $w_{10}(\ell,L) = A
\ell/(\ell^{1.76}(L-\ell)^{1.76})$.  Note that this functional form is
still singular at $\ell\to0$, although the singularity is weakened
compared to the case without the sliding ring.  This singularity is
not visible in Fig.~\ref{fig:fig1}, presumably due to the short size
of the simulated chains.

\begin{figure}
\centerline{\includegraphics[height=1.5in]{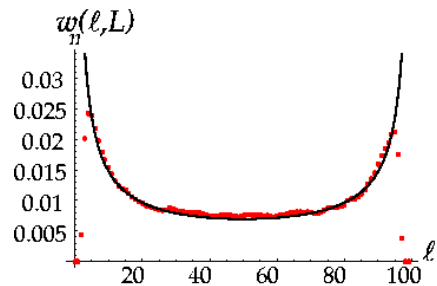}}
\caption{Probability distribution of length for two loops with one ring on each
side ($L=100$). }
\label{fig:1and1ring}
\end{figure}

\begin{figure}
\centerline{\includegraphics[height=1.5in]{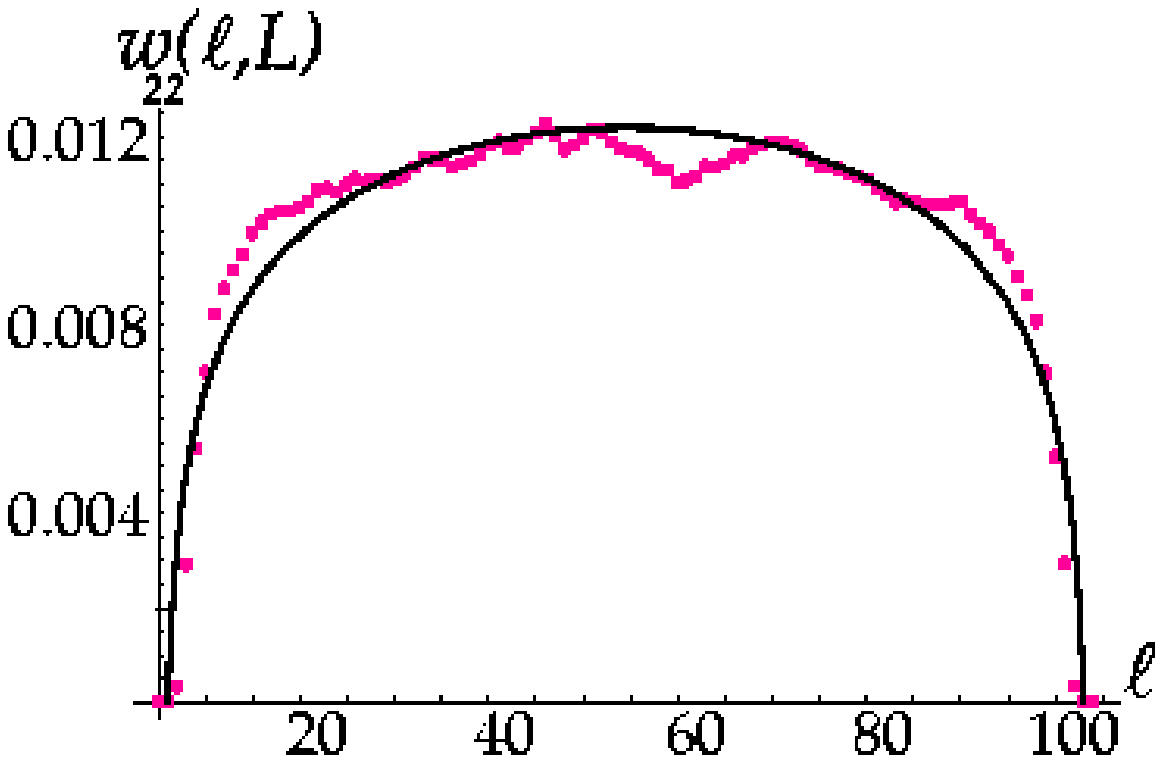}}
\caption{Probability distribution for two loops with two rings on
each side ($L=100$). }
\label{fig:2and2ring}
\end{figure}

\begin{figure}
\centerline{\includegraphics[height=1.5in]{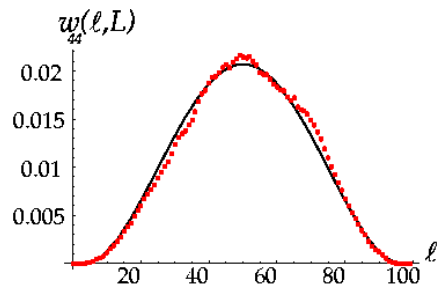}}
\caption{Probability distribution for two loops with four rings on
each side ($L=100$).}
\label{fig:4and4ring}
\end{figure}

The entropic effects can be increased even further by placing more
sliding rings on the loops of Fig.~\ref{fig:fig1}. The presence of
$n$ sliding-rings on a loop enhances the number of configurations
to
\begin{eqnarray}
\label{slidingring}
w_n \sim \frac{\ell !} {(\ell -n )! n!}  w_{0}\, \, .
\end{eqnarray}
The side with the larger number of sliding rings is expected to
dominate in a competition. Figure \ref{fig:1and1ring} shows the
probability distribution when we have one ring on {\em each} side.
The solid line in Fig.~\ref{fig:1and1ring} corresponds to the
formula,
\begin{eqnarray}
\label{dist1} w_{nn}(\ell,L) =A \frac{\ell!  (L-\ell)!} {(\ell -
n)!  (L-\ell -n)! \ell^{1.76}(L-\ell)^{1.76}}\, \, ,
\end{eqnarray}
with $n=1$.  The dots on the figures are the result of a simulation of
$10^{9}$ MC steps.  Similar results have been observed for chains with
sizes $L=50$, $150$, and $200$.  The corresponding results with two
rings on each side are shown in Fig.~\ref{fig:2and2ring}; the solid
line represents Eq.~(\ref{dist1}) with $n=2$.  Note the dramatic
difference between the shapes of these two figures: with one ring on
each side the distribution is peaked (in fact singular) at the two
extremes, while for $n=2$, the maximum has moved to the center.  This
trend becomes even more pronounced in Fig.~\ref{fig:4and4ring} which
illustrates the case with $n=4$.  Clearly, increasing the number of
rings results in a probability distribution which peaks more sharply
in the middle.  The solid line in Fig.~\ref{fig:4and4ring} again
represents Eq.~(\ref{dist1}) with $n=4$.  The good matches between the
probability distributions obtained in simulations, and
Eq.~(\ref{dist1}) for different numbers of rings $n$ lend further
credence to the validity of our method.

\section{ Knots}\label{sec:knot}

\subsection {Background} \label{subsec:back}

We are now in position to apply the entropic competition method to the
more complicated problem of knots.  Knots frequently appear in closed
polymers and play a major role in numerous biological systems.  For
example, during transcription of DNA, a variety of ``de-knotting''
enzymes, called topoisomerases, remove knots and entanglements to
allow this process to go forward\cite{alberts,drew}.  Understanding of
the action of these enzymes has been improved with the help of knot
theory.  Knotted configurations have also been observed in some
proteins and interfere with their folding into the proper shapes and
thus lead to diseases, such as Alzheimer's and primary
amyloidosis\cite{fink}.  While knotted configurations hamper the
proper functioning of a number of biological and chemical systems,
they have crucial and positive roles in many others.  The tertiary
structure of RNA is an example in which the topological entanglements
may have positive influences.  There has been extensive research on
predicting and understanding the relation between structure and
function in RNA \cite{bust}.  The folded structure of some types of
RNA molecules (Ribozymes) determine their catalytic activities which
are crucial to the functioning of the cell\cite{alberts}. 
Pseudo-knots, which are formed by base pairing between a loop and some
region outside the loop, have been found in various kinds of RNAs and
are recognized as simple RNA folding motifs\cite{batenburg}.  A more
physical example is provided by permanent entanglements in rubber,
which restrict the number of allowed configurations of each segment,
and thus influence the elasticity of rubber\cite{rubber}.

On the experimental front, artificial knots have been tied on both
DNA and actin filaments \cite{arai}, and the first Borromean DNA
rings have also been assembled\cite{mao}.  Single molecule
techniques have provided a powerful tool to examine a wide variety
of physical properties of knotted polymers \cite{rybenkov}.
Despite all this progress, our knowledge about the typical
conformations and physical properties of knotted polymers remains
rudimentary.  
This is due to the difficulty of incorporating topological constraints 
in the analytical treatments of the statistics of polymers, 
as mathematical methods of knot detection are mainly of
``algorithmic'' nature.\cite{kauf}. 
This
complexity has encouraged the use of Monte Carlo simulations, such
as the one reported here.

\subsection{Competing knots}

How does the altered topology of a closed curve with a knot modify
the number of available configurations?  To investigate this
issue, Orlandini {\em et al.}\cite{orlandini} performed grand
canonical Monte Carlo simulations of SA polygons with a fixed knot
type $\cal{K}$.  In these simulations the length of the chain is
not fixed, but the set of allowed moves is such that the topology
of the chain is preserved. Orlandini {\em et al.} conjectured that
the number of configurations of knotted loops takes the same
general form as Eq.~(\ref{w0}), asymptotically behaving as
\begin{equation}
 w_{\cal K} \sim A_{\cal K}\mu_{\cal K}^{\ell} \ell^{\alpha({\cal
 K)}-3}, \label{wk}
\end{equation}
with parameters that may depend on the knot type ${\cal K}$.  Assuming
this form, one can find an expression for the mean length
$\left\langle n({\cal K})\right\rangle$, a quantity that can be
measured in the grand canonical simulations.  Fitting the simulation
data for $\left\langle n({\cal K})\right\rangle^{-1}$, and
extrapolating the results to the limit of infinite size, Orlandini
{\em et al.} can estimate the parameters in Eq.~(\ref{wk}).  In
particular they confirm that the effective growth constant $\mu_{\cal
K}$ is independent of the knot type.  Assuming that this is the case,
they then conclude that the universal power-law exponents behave as
$\alpha({\cal K}) = \alpha(\emptyset)+N_{f}$, where $\emptyset$ refers
to a simple loop (unknot), while $N_{f}$ is the number of prime
factors in the knot type $\cal K$.  Such a conclusion has a simple and
elegant interpretation: each prime factor of the knot becomes a tight
element that incorporates an asymptotically small fraction of the
monomers.  The tight factors can occupy any position along the
remaining large loop, each increasing the number of configurations by
a factor of roughly the size of the loop, much like the sliding rings
discussed in the previous section.
\begin{figure}
\centerline{\includegraphics[height=1.5in]{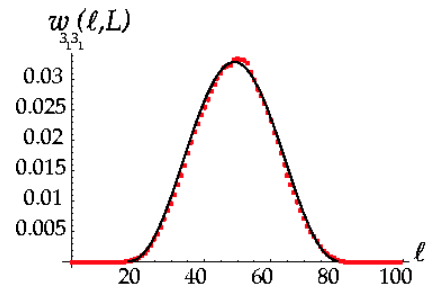}}
\caption{Probability distribution for two loops with one knot on each
side ($L=100$).}
\label{fig:22knots}
\end{figure}

If a prime knot, such as a trefoil, increases the number of
configurations of the closed polymer by the same factor as a sliding
ring, the two should behave similarly in the arena of ``entropic
competition."  Indeed a knotted loop added to one side of
Fig.~\ref{fig:fig1} pulls the entire chain on its side as in
Fig.~\ref{fig:ring1}.  However, when we place one trefoil knot on each
of the loops, the resulting segment distribution, as plotted in
Fig.~\ref{fig:22knots} is qualitatively different from the curve found
when a sliding ring appears on each side (Fig.~\ref{fig:1and1ring}). 
The distribution for the competing knots is peaked at the center, much
like cases with more than one sliding ring on each side.  The solid
line in Fig.~\ref{fig:1and1ring} in fact represents Eq.~(\ref{dist1})
with $n=4$.  We do not suggest that the trefoil is asymptotically
similar to four rings, but that this is a good effective description
of a trefoil knot with a few hundred monomers.  To obtain more insight
on finite-size effects, we compare the probability distributions of
knotted loops with different sizes ranging from $L=50$ to $L=200$.

\begin{figure}
\centerline{\includegraphics[height=1.5in]{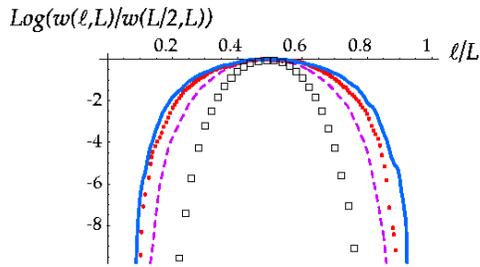}}
\caption{A
logarithmic plot of probability density for two loops with one
knot on each side for $N=50$ (square), 100 (dashed line), 150
(dots), and 200 (solid line).  The graphs were vertically displaced
to coincide at $\ell/L=0.5$. As $L$ increases, the distributions
become wider.} \label{fig:logrescale}
\end{figure}

Figure ~\ref{fig:logrescale} shows the logarithmic plot of rescaled
probability densities for $N=50$, 100, 150, and 200, as a function of
$\ell/L$.  The $y$ coordinates are shifted so that the maxima of
distributions for all sizes coincide.  We note that the maximum
becomes flatter upon increasing $L$.  The lack of data collapse is a
clear indication of finite-size effects.
To check this, we performed similar simulations with four rings on
each side with chains of lengths $L=50$ through $L=200$.  In contrast
to knots (Fig.~\ref{fig:logrescale}), we observed that all rescaled
curves collapsed in this case.

\section{Knots versus Rings} \label{sec:compete}

\begin{figure}
\centerline{\includegraphics[height=1.5in]{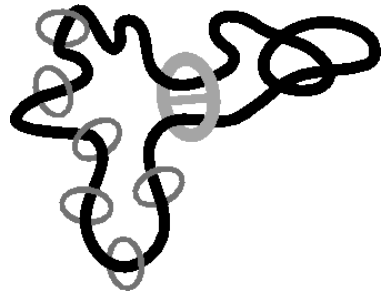}}
\caption{Schematic ``entropic competition'' between a trefoil knot and 6 sliding rings.}
\label{fig:competition}
\end{figure}

In order to compare the relative ``strengths'' of knots and rings in
increasing the number of configurations, we performed several
simulations in which we pitted a trefoil knot against different
numbers of rings, as depicted in Fig.~\ref{fig:competition}.  While
the hypothesis of a tight knot\cite{orlandini,kat,ralf} suggests that
the trefoil should be well matched to a single ring, in our
simulations with chains of 100 monomers, we find that around 6 rings
are necessary to prevent the chain from being pulled completely to the
side with the trefoil.  Part of this effect is no doubt due to the
size of the knot: even in its most compact form the trefoil knot
involves around 14 monomers, while the sliding ring is assumed to
occupy only one.  In actuality, even a `tight knot' will most likely
be considerably bigger than the minimal size, with typical sizes that
grow with the length of the chain\cite{farago}.  To test for these
effects we performed some further studies as reported in this section.

To take account of the minimal size of the trefoil knot, we also added
a similar constraint to the side with sliding rings.  In these
simulations the monomers were prevented to move from the ring side to
the knot side with a probability of $1/(\ell-13)$.  This means that
when $\ell =14$, the monomer is strictly forbidden to pass through the
hole from the ring side to the knot side; thus, at least 14 monomers
remain on each side throughout the simulation.  Even in this case, we
observe that the trefoil ``wins'' the competition against a single
sliding ring.  Once more, we found that at least four rings are need
for the two sides of the chain to exert equal amount of force on each
other.  Thus the minimal size of the knot is not a crucial issue.

\begin{figure}
\centerline{\includegraphics[height=1.5in]{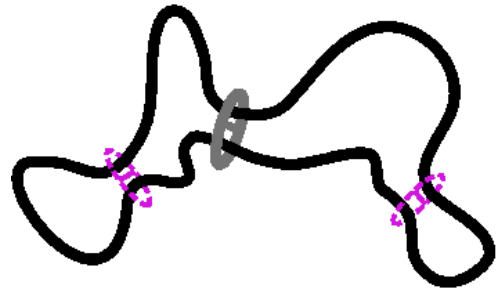}}
\caption{Two loops are separated by the thick oval shaped hole.
The dotted ovals are slip-links.}
\label{fig:sliplinks}
\end{figure}

We certainly do not expect the trefoil to be maximally tight.  The
number of monomers participating in the knot region is itself a
not-so-well-defined and fluctuating quantity.  A better model than the
sliding ring, which allows for this possibility, is the slip-linked
para-knot depicted in Fig.~\ref{fig:sliplinks}.  The thick oval ring
in this figure represents the `worm-hole' separating the two segments,
while the dotted ovals are slip-links which keep two points on each
loop close to each other (creating a figure-8 structure on each
segment).  The effect of each slip-link is similar to the hole in that
they create two loops; however, there are self-avoiding interactions
among all the monomers of a loop made into a figure-8 by the
slip-link.  The number of configurations of a figure-8 structure with
self-avoiding constraints for $1 \ll \ell \ll L$ scales as
$\ell^{-c}(L-\ell)^{-c}$ with $c=2.88$ in $d=2$ and $c=2.24$ in $d=3$
\cite{duplantier}.  Although this formula is strictly valid only for
large $\ell$ and $L$, a recent experiment on a figure-8 chain on a
vibrating plate in 2-dimensions is in good agreement with this
formula\cite{ben,note1}.

\begin{figure}
\centerline{\includegraphics[height=1.25in]{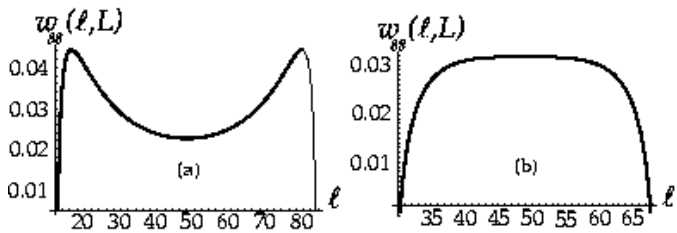}}
\caption{ Probability distribution for two loops with one slip-link on
each side (figure-8), from Eq.~(\ref{paraknot}) with $L=100$ and $s_{min}=14$ (a)
and $s_{min}=30$ (b).}
\label{fig:slip100}
\end{figure}

Extending the asymptotic formula for the figure 8 to all separations,
we can obtain an analytic form for the number of configurations of
two figure 8's in competition.
Assuming a minimal size $s_{min}$ for each segment, the number
of configurations is
\begin{eqnarray}
\label{paraknot}
w_{8,8}(\ell,L)\simeq\int_{s_{min}}^{\ell} ds \frac{(\ell-s)}
{s^{c}(\ell-s)^{c}} \times \int_{s_{min}}^{L-\ell} ds
\frac{(L-\ell-s)} {s^{c}(L-\ell-s)^{c}} \, \, .
\end{eqnarray}
Each integrand represents the probability density of a loop with its
associated para-knot.  Since there exists self-avoiding interaction
between each loop and its para-knot, we set $c=2.88$ and $c=2.24$ for
two and three dimension respectively.  Figure~\ref{fig:slip100}a is a
graph of Eq.~(\ref{paraknot}) with $s_{min}=14$ and $c=2.24$.  The
distribution shown in Fig.~\ref{fig:slip100}a is qualitatively similar
to the one in Fig.~\ref{fig:1and1ring}, for two loops with {\em one}
sliding-ring on each side, in that for both cases, the minimum of the
distributions is in the middle and their maxima are on the sides.  If
we increase the minimal size from $s_{min}=14$ to $s_{min}=30$, the
maximum moves to the middle as depicted in Fig.~\ref{fig:slip100}b.

\section{  Conclusions}

An important quantity in polymer physics is the number of
configurations that a chain of length $\ell$ can take.  As noted
in the introduction, this quantity has the asymptotic form of
$\mu^\ell \ell^{-c}$ where $\mu$ depends on the microscopic
features of a chain while the exponent $c$ is universal. 
In this paper we focus on obtaining the universal exponent,
while bypassing the parameter $\mu$.
To this end, we employ a hole (or buckle) to divide a polymer into two
segments (see Figs.~\ref{fig:all},~\ref{fig:fig1}), and then allow
the two parts to exchange monomers and to ``entropically''
compete.  The resulting histograms allow us to calculate the
entropic exponents with a variety of boundary conditions and/or
topological constraints.

The scaling form in Eq.  ~\ref{length} is only valid
asymptotically, and there are in general corrections to
scaling which complicate the extraction of entropic exponents for
short polymers.  In the application of entropic competition to
knots, we find that finite size effects are so pervasive that we
cannot extract reliable universal information pertaining to knots
for ($L<300$).  (We find an effective exponent of around 4 for a
trefoil knot over the simulated lengths.)  Indeed,
simulations performed by Farago {\em et al.}\cite{farago} indicate
that while the ratio of the knot size $N_{\cal K}$ to the total
length of the chain decreases, the effective knot size grows with
the size of the polymer, suggesting that simulations on very large
chains are needed to avoid the finite size effect. An important
question which remains is at what size of a molecule the universal
predictions of the asymptotic theories are expected to hold.

The primary goal of this paper is to present the method of ``entropic
competition'' for calculation of entropic exponents of polymers with
different boundary conditions and under different topological
constraints.  Our procedure for extracting exponents is in principle
simple, yet capable of producing accurate results.  An intriguing
application of the method to topologically constrained chains, such as
knotted polymers, is also attempted in this paper.  This example also
illustrates the limitations of the technique, in that we find that
finite-size effects are quite important to our simulated knotted
chains with $L<300$ monomers.

It's noteworthy to mention that each simulation reported in this
paper was obtained in a maximum of two week's CPU time on a
desk-top computer.  One can apply ``entropic competition'' method
to longer chains and calculate ``entropic exponents'' with higher
accuracy on a large cluster of computers with longer amounts of
CPU time.

\section{Acknowledgements}

The authors would like to acknowledge helpful discussions with J.
Rudnick, M. Deserno, and R. Metzler.  This work was supported by NSF
grant DMR-01-18213, and Israel Science Foundation Grant 38/02.  RZ
acknowledges support from the UC President's Postdoctoral Fellowship
program.


\begin{thebibliography}{000}
\bibitem{note1} The experiment in Ref.  \cite{ben} is done with a
vibrated granular chain far from equilibrium.  Nonetheless, their
results are consistent with the theoretical equilibrium predictions.
\end{thebibliography}
\end{document}